\documentclass[preprint,12pt,authoryear]{elsarticle}
\setcitestyle{numbers}
\biboptions{square,numbers,sort&compress}
\usepackage{amssymb}
\usepackage{amsmath}
\usepackage[linesnumbered,ruled]{algorithm2e}

\newtheorem{definition}{Definition}
\newtheorem{theorem}{Theorem}

\newtheorem{example}{Example}

\newtheorem{remark}{Remark}

\newtheorem{proposition}{Proposition}

\begin{document}

\begin{frontmatter}

\title{Non-interference analysis of bounded labeled Petri nets} 

\author[aff1]{Ning Ran}
\author[aff2]{Zhengguang Wu}
\author[aff3]{Shaokang Zhang}
\author[aff4]{Zhou He}
\author[aff5]{Carla Seatzu}

\address[aff1]{Laboratory of Energy-Saving Technology, College of Electronic \& Information Engineering, Hebei University, Baoding 071002, China. {\tt\small ranning87@hotmail.com}}
\address[aff2]{College of Control Science and Engineering, Zhejiang University, Hangzhou 310027, China. {\tt\small nashwzhg@126.com}}
\address[aff3]{School of Cyber Security and Computer, Hebei University, Baoding 071002, China. {\tt\small zhangshaokang@hbu.edu.cn}}
\address[aff4]{College of Mechanical \& Electrical Engineering, Shaanxi University of Science \& Technology, Xi'an 710021, China. {\tt\small hezhou@sust.edu.cn}}
\address[aff5]{Department of Electrical and Electronic Engineering, University of Cagliari, Cagliari 09124, Italy. {\tt\small carla.seatzu@unica.it}}

\begin{abstract}
This paper focuses on a fundamental problem on information security of bounded labeled Petri nets: non-interference analysis.
As in hierarchical control, we assume that a system is observed by users at different levels, namely high-level users and low-level users.
The output events produced by the firing of transitions are also partitioned into high-level output events and low-level output events.
In general, high-level users can observe the occurrence of all the output events, while low-level users can only observe the occurrence of low-level output events.
A system is said to be non-interferent if low-level users cannot infer the firing of transitions labeled with high-level output events by looking at low-level outputs.
In this paper, we study a particular non-interference property, namely strong non-deterministic non-interference (SNNI), using a special automaton called SNNI Verifier, and propose a necessary and sufficient condition for SNNI.
\end{abstract}


\end{frontmatter}

\section{Introduction}\label{section1}

Information security is quite important for nowadays control systems.
In the past few decades, some problems related to information security of discrete event systems (DESs), such as \textit{opacity}~\cite{01,04,05,06,07,08,09,10,11}, \textit{non-interference}~\cite{12,13,14,15,16,17,18,19,22,23}, \textit{anonymity}~\cite{20,21}, were extensively studied and a series of methods have been developed.

Information leak is a typical security problem and has received a lot of attention.
In such a scenario, an intruder may speculate sensitive information based on the released data.
In the DES framework, this problem can be formalized as a \textit{non-interference problem} such as \textit{strong nondeterministic non-interference (SNNI)} and \textit{bisimulation strong non-deterministic non-interference (BSNNI)}.

In~\cite{12}, the formal definition of SNNI was first proposed for automata, where the system is assumed to be monitored by high-level users (e.g. the administrators) and low-level users (e.g. the customers) from different points of view.
Although both classes of users know the structure of the system, the high-level users can observe all the events, while the low-level users can only observe a subset of events, namely the set of low-level events.
A system is said to be SNNI if for any enabled sequence that contains high-level events, its projection on the set of low-level events is still enabled.
In other words, if a system is SNNI, then low-level users cannot infer the occurrence of high-level events no matter which low-level events are observed.
In addition, some related security properties of automata are also studied and compared with SNNI in~\cite{12}.

In~\cite{13}, the problem of SNNI was first rephrased in the Petri net framework.
Two new non-interference properties are defined taking advantage from the structural feature of safe Petri nets.
Finally, reference~\cite{13} proves that the absence of causal places and conflict places is a sufficient condition for the structural non-interference.
Best \textit{et al.}~\cite{14} prove that the property of SNNI and the property of \textit{nondeducibility on composition (NDC)} are equivalent, and also prove that the latter is decidable for unbounded Petri nets.
Basile \textit{et al.}~\cite{16} investigate SNNI of Petri nets based on the solution of some integer linear programming (ILP) problems, and propose a necessary and sufficient condition for SNNI.
The property of BSNNI is also studied in this paper.
This method is then improved by removing the assumption of acyclicity of Petri nets~\cite{17}.
To enforce SNNI to Petri nets, an online supervisory control policy is proposed in~\cite{18} that dynamically disables a subset of transitions whose occurrence may violate the conditions for SNNI.
In~\cite{19}, Basile~\textit{et al.} extend such ILP-based methods to the SNNI and BSNNI analysis of labeled Petri nets.
Although the assumption that the low-level subnet is acyclic is not required in these papers, authors use the linear characterizations of the firing vectors enabled at a certain marking with a number of constraints that linearly increases with the length of the corresponding sequence.

To avoid exhaustive enumeration of the whole state space, Ran \textit{et al.}~\cite{22} recently propose new methods for SNNI analysis and
enforcement of bounded Petri nets based on the notion of basis marking.
They prove that the property of SNNI can be analyzed checking the set of arcs in the basis reachability graph (BRG).
To enforce SNNI, a supervisory control policy is designed that disables a subset of low-level transitions as a response to each transition sequence generated by the system.
Such methods have also been extended to the analysis and enforcement of BSNNI of Petri nets in~\cite{23}.

The method proposed in~\cite{22} cannot be directly applied to labeled Petri nets, in which two or more transitions that are simultaneously enabled may produce the same output event.
In such a case, the problems of SNNI analysis and enforcement become more complicated since to a low-level transition sequence whose enabling is due to the occurrence of high-level transitions, there may correspond another low-level transition sequence sharing the same output event sequence.
This paper extends the analysis method in~\cite{22} to the case of labeled Petri nets.
More precisely, we first construct an unfolded version of the BRG to identify all the low-level transition sequences whose enabling may only be due to the occurrence of high-level transitions.
Then a special automaton, called \textit{SNNI Verifier}, is introduced to check whether every such sequence is indistinguishable from another low-level transition sequence whose enabling is not due to the occurrence of high-level transitions.
Based on such a Verifier, a necessary and sufficient condition is given for SNNI.
Overall, the main contributions of this paper may be summarized as follows:
\begin{itemize}
  \item We formalize the relationship between the language of the BRG and the SNNI property of labeled Petri nets. Then, the notion of \textit{unfolded basis reachability graph (UBRG)} is proposed, which facilitates us to identify the set of sequences that are related to SNNI.
  \item We define a special automaton called SNNI Verifier, and give a necessary and sufficient condition for SNNI of labeled Petri nets.
\end{itemize}

The rest of the paper is organized as follows.
Section~\ref{section2} provides the required preliminaries on labeled Petri nets.
In Section~\ref{section3}, the notions of basis marking and BRG are first recalled, then the notion of UBRG is presented.
We introduce the notion of SNNI Verifier and propose a necessary and sufficient condition for SNNI in Section~\ref{section4}.
Finally, conclusions are drawn in Section~\ref{section6}, where our future research directions in this framework are also illustrated.

\section{Preliminaries}\label{section2}

\subsection{Petri Nets}\label{section2.1}

A Petri net (PN) is a 4-tuple $N=(P,T,F,W)$, where $P$ is the set of places, $T$ is the set of transitions, $F\subseteq(P\times T)\cup(T\times P)$ is the set of arcs, $W$ is a mapping that assigns a non-negative integer weight to an arc: $W(x,y)>0$ if $(x,y)\in F$, and $W(x,y)=0$  if $(x,y)\notin F$, where $x, y\in P\cup T$.
The incidence matrix $[N]$ is a $|P|\times |T|$ integer matrix with $[N](p,t)=W(t,p)-W(p,t)$, where $p\in P$ and $t\in T$.
Given a place $p$ (transition $t$), its incidence vector, a row (column) in $[N]$, is denoted
by $[N](p,\cdot)$ ($[N](\cdot,t)$).
Let $x\in P\cup T$ be a node of net $N$.
The preset of $x$ is $^\bullet x=\{y\in P\cup T \mid (y,x)\in F$\} while the postset is $x^\bullet=\{y\in P\cup T \mid (x,y)\in F$\}.

A marking $m$ of a PN $N$ is a mapping from $P$ to $\mathbb{N}=\{0,1,2,...\}$; $m(p)$ denotes the number of tokens in $p$.
$(N,m_0)$ denotes a PN system with initial marking $m_0$.

A transition $t$ is enabled at marking $m$ if $\forall p\in{^\bullet t}, m(p)\geq W(p,t)$, which is denoted by $m[t \rangle$.
Firing $t$ at $m$ yields a new marking $m'$ such that $\forall p\in P, m'(p)=m(p)+[N](p,t)$, which is denoted by $m[t\rangle m'$.
The notation $m[s\rangle$ denotes that the transition sequence $s\in T^*$ is enabled at $m$.
Marking $m''$ is said to be reachable from $m$ if there exists a transition sequence $s$ such that $m[s\rangle m''$.
The set of markings reachable from $m$ in $N$ is called the reachability set of $(N,m)$ and is denoted by $R(N,m)= \{m'\in \mathbb{N}^{|p|} \mid \exists s\in T^*, m[s \rangle m'\}$.
The set of all transition sequences that are enabled at the initial marking $m_0$ is $\mathcal{L}(N,m_0)=\{s\in T^* \mid m_0[s\rangle\}$.

Given a transition sequence $s\in T^*$, we denote by $\pi(s)$ the Parikh vector of $s$.
Let $y=\pi(s)$, we write $y(t)=k$ if transition $t$ appears $k$ times in $s$.

A PN is said to be \textit{bounded} if there exists a positive constant $k$ such that $\forall p\in P$, $\forall m\in R(N,m_0)$, $m(p)\leq k$.
It is \textit{unbounded} if it is not bounded.
A sequence of nodes $x_1x_2\ldots x_n$ is called a path if $x_{i+1}\in x_i^\bullet$, where $x_i\in P\cup T$.
If the first node is the same as the last node, i.e., $x_1=x_n$, then the path is called a \textit{circuit}.
A PN with no circuits is said to be \textit{acyclic}.

Let $T'\subseteq T$ be a subset of transitions, the subnet $N'=(P,T',F',W')$ is called the \textit{$T'$-induced subnet} of $N$, where $F'$ is the restriction of $F$ to $(P\times T')\cup(T'\times P)$ and $W'$ is the restriction of $W$ to $(P\times T')\cup(T'\times P)$.

\subsection{Labeled Petri Nets}

A labeled PN system (LPNS) is a 4-tuple $(N,m_0,l,A)$, where $A$ is the alphabet of labels and $l: T^*\rightarrow A^*$ is a labeling function defined as follows:
\begin{itemize}
  \item $l(t) \in A$, if $t\in T$;
  \item $l(\varepsilon)=\varepsilon$, where $\varepsilon$ is the empty string;
  \item $l(st)=l(s)l(t)$, if $s\in T^*$ and $t\in T$.
\end{itemize}
When a transition fires it produces an output event which corresponds to the label associated with it.
In simple words, the label can be seen as the output event produced by a sensor associated with the event modeled by the corresponding transition.

Transitions are called \textit{indistinguishable} if they share the same label (or equivalently, the same output event); alternatively, they are called \textit{distinguishable}.
Given a label (or output) sequence $w\in A^*$, we denote by $l^{-1}(w)$ the set of transition sequences that are consistent with $w$, namely $$l^{-1}(w)=\{s\in \mathcal{L}(N,m_0) \mid l(s)=w\}.$$
The language of labels is denoted by $$l(\mathcal{L}(N,m_0))=\{l(s)\in A^*\mid s\in\mathcal{L}(N,m_0)\}.$$

The alphabet $A$ can be partitioned as $A=A_E\cup A_I$, where $A_E$ denotes the set of \textit{explicit} labels, $A_I$ denotes the set of \textit{implicit} labels and $A_E\cap A_I = \emptyset$.
Accordingly, the set $T$ of transitions is partitioned as $T=T_E\cup T_I$, where $T_E$ is the set of explicit transitions, i.e.,
$$T_E = \{t\in T\mid l(t) \in A_E\};$$
$T_I$ is the set of implicit transitions, i.e.,
$$T_I = \{t\in T\mid l(t) \in A_I\}.$$
We denote by $[N]_E$ ($[N]_{I}$) the restriction of $[N]$ to $T_E$ ($T_{I}$), namely, $[N]_E$ is a $|P|\times |T_E|$ integer matrix with $[N]_E(p,t)=W(t,p)-W(p,t)$, where $p\in P$ and $t\in T_E$.

The projection $P_E:T^*\rightarrow T_E^*$ is defined as:
\begin{itemize}
  \item $P_E(\varepsilon)=\varepsilon$;
  \item for all $s\in T^*$ and $t\in T$, $P_E(st)=P_E(s)t$ if $t\in T_E$, and $P_E(st)=P_E(s)$ otherwise.
\end{itemize}
Analogously, we define the projection $P_{I}:T^*\rightarrow T_{I}^*$.
The projections of $\mathcal{L}(N,m_0)$ over $T_E$ and $T_{I}$ are denoted by $P_E(\mathcal{L}(N,m_0))$ and $P_{I}(\mathcal{L}(N,m_0))$, respectively.

\subsection{Nondeterministic Finite Automata}

A nondeterministic finite automaton (NFA) is a 4-tuple $G=(\Theta,E,\Delta,X_0)$, where $\Theta$ is a finite set of states, $E$ is a set of events, $\Delta\subseteq\Theta\times E\times\Theta$ is the transition relation, $X_0\subseteq\Theta$ is a set of initial states.
An NFA with labels is a 6-tuple $(\Theta,E,\Delta,X_0,l,A)$, where $A$ is the alphabet of labels and $l: E^*\rightarrow A^*$ is a labeling function that is similar to the labeling function defined in LPNs.

\section{Unfolded Basis Reachability Graph}\label{section3}

In this section, we first recall the notions of basis marking and basis reachability graph (BRG), then introduce the notion of \textit{unfolded BRG} (UBRG).

\subsection{Basis Reachability Graph}

The notion of basis marking is first proposed in~\cite{24} for estimating the markings of PNs in the presence of silent transitions.
It provides a compact way to describe the set of reachable markings consistent with a given observation.
Then this method has been efficiently extended to study some problems related to partial observation, such as fault diagnosis~\cite{25,26,27}, fault prognosis~\cite{29}, reachability~\cite{31}, etc.
We adopt the following two assumptions, which are necessary to apply the method based on the notion of basis marking.
\begin{itemize}
  \item[A1)] The PN system is bounded.
  \item[A2)] The $T_{I}$-induced subnet is acyclic.
\end{itemize}

Note that $T_I$-induced subnet is a subnet $(P,T_I,F_I,W_I)$, where $F_I$ is the restriction of $F$ to $(P\times T_I)\cup(T_I\times P)$ and $W_I$ is the restriction of $W$ to $(P\times W_I)\cup(W_I\times P)$.
Although Assumption A2 limits the application scope of the proposed method, it is reasonable for a real system since otherwise the system may fall into an infinite loop without generating any detectable events.

Given a marking $m \in R(N,m_0)$ and a transition $t\in T_E$, the set of \textit{explanations} of $t$ at $m$ is defined as $$\Sigma(m,t)=\{s\in T_{I}^* \mid m[s\rangle m', m'[t\rangle\},$$
and the set of \textit{e-vectors} is defined as $$Y(m,t)=\{\pi(s) \mid s\in\Sigma(m,t))\}.$$

The set of \textit{minimal explanations} of $t$ at $m$ is defined as
$$\Sigma_{min}(m,t)=\{s\in\Sigma(m,t) \mid \nexists s'\in\Sigma(m,t):\pi(s')\lneq\pi(s)\}.$$
The set of \textit{minimal e-vectors} is denoted by $$Y_{min}(m,t)=\{\pi(s) \mid s\in\Sigma_{min}(m,t))\}.$$
In particular, $Y_{min}(m,t)$ can be computed by Algorithm~4.4 in~\cite{26}.

In simple words, $\Sigma(m,t)$ is the set of implicit sequences whose firing at $M$ enables $t$.
Among the above sequences, $\Sigma_{min}(m,t)$ selects those whose Parikh vector is minimal.
The Parikh vectors of these sequences are called minimal e-vectors.

Let $(N,m_0,l,A)$ be an LPNS and $w\in A_E^*$ be a sequence of explicit labels. 
The set of pairs ($s_E\in T_E^*$ with $l(s_E)=w$ and the \textit{justification}) is indicated as
$$\hat{\mathcal{J}}(w)=\{(s_E,s_I),s_E\in T_E^*,l(s_E)=w,s_I\in T_I^* \mid [\exists s\in l^{-1}(w):s_E=P_E(s),s_I=P_I(s)]$$
$$\wedge[\nexists s'\in l^{-1}(w):s_E=P_E(s'),s_I'=P_I(s') \wedge\pi(s_I')\lneq\pi(s_I)]\},$$

\noindent and the set of pairs ($s_E\in T_E^*$ with $l(s_E)=w$ and the j-vector) is denoted by
$$\hat{Y}_{min}(m_0,w)=\{(s_E,y),s_E\in T_E^*,l(s_E)=w,y\in\mathbb{N}^{|T_I|} \mid \exists(s_E,s_I)\in\mathcal{\hat{J}}(w):\pi(s_I)=y\}.$$

In other words, $\hat{\mathcal{J}}(w)$ is the set of pairs whose first element is the sequence $s_E$ labeled $w$ and whose second element is the corresponding sequence of implicit transitions interleaved
with $s_E$ whose firing enables $s_E$ and whose Parikh vector is minimal.
The Parikh vectors of these implicit sequences are called j-vectors (justification vector).
$\hat{Y}_{min}(m_0,w)$ is the set of pairs whose first element is the sequence $s_E$ labeled $w$ and whose second element is the corresponding j-vector.

The set of \textit{basis markings} of $w$ is indicated as
$${M}_b(w)=\{m\in \mathbb{N}^{|P|} \mid m=m_0+[N]_I\cdot\pi(s_I)+[N]_E\cdot\pi(s_E), (s_E,s_I)\in\mathcal{\hat{J}}(w)\},$$
\noindent and $\mathcal{M}_b=\bigcup\limits_{w\in l(\mathcal{L}(N,m_0))}{M}_b(w)$.

Therefore, a basis marking is a marking that can be reached from the initial marking firing a sequence of transitions that is consistent with the sequence of explicit labels and a sequence of implicit transitions, interleaved with the previous sequence, whose firing is strictly necessary to enable it (in the sense that its Parikh vector is minimal)~\cite{26}.
The set of basis markings is a subset (usually a strict subset) of the set of reachable markings. Therefore, if the net is bounded, the set of basis markings is finite.

\begin{definition}[\cite{25}]
Let $(N,m_0,l,A)$ be an LPNS, the \textit{BRG} $G=(\Theta,E,\Delta,m_0)$ is an NFA, where:
\begin{itemize}
  \item $\Theta=\mathcal{M}_b$;
  \item $E\subseteq(T_E \times \mathbb{N}^{|T_{I}|})$ is the set of events;
  \item $\Delta\subseteq \Theta\times E\times \Theta$ is the transition relation;
  \item $m_0$ is the initial state.
\end{itemize}
The BRG is constructed by Algorithm 1.\label{def1}
\end{definition}

\begin{algorithm}
\caption{BRG Construction}\label{a1}
\KwIn{An LPNS $(N,m_0,l,A)$, where $A=A_E\cup A_I$.}
\KwOut{The BRG $G=(\Theta,E,\Delta,m_0)$.}

Let $m_0$ be the root node, $\Theta = \{m_0\}$, $E = \emptyset$, $\Delta = \emptyset$.\\
\While{nodes with no tag exist,}
{
  choose a node $m\in\Theta$ with no tag,\\
  \For{all $t\in T_E$,}
  {
    \If{$Y_{min}(m,t)\neq\emptyset$,}
    {
      \For{all $y\in Y_{min}(m,t)$,}
      {
        let $m' = m+[N]_{I}\cdot y+[N](\cdot,t)$,\\
        let $\Theta = \Theta \cup \{m'\}$, $E = E\cup \{(t,y)\}$ and $\Delta = \Delta \cup \{(m,(t,y),m')\}$.
      }
    }
  }
  tag node $m$ ``old''.
}
Remove all tags.
\end{algorithm}

From Algorithm 1, we know that for each transition relation $(m,(t,y),m') \in \Delta$, $y$ is a minimal e-vector of $t$ at $m$ and $m' = m+[N]_{I}\cdot y+[N](\cdot,t)$.

Note that the BRG constructed by Algorithm 1 is slightly different from the BRG proposed in~\cite{25}.
In the BRG constructed in~\cite{25}, each node contains two entries, where the second entry is a binary value that allows us to keep track of some information which is not relevant now, so it is omitted.

\begin{figure}[!t]
  \centering
  \includegraphics[scale=0.9]{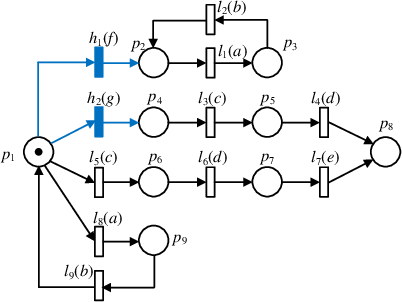}\\
  \caption{A labeled Petri net system.}\label{fig1}
\end{figure}

\begin{figure}[!t]
  \centering
  \includegraphics[scale=1]{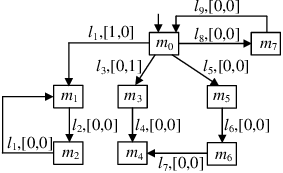}\\
  \caption{BRG $G$ of the LPNS in Fig.~\ref{fig1}.}\label{fig2}
\end{figure}

\begin{example}
Consider the LPNS in Fig.~\ref{fig1}, where $A_E=\{a,b,c,d,e\}$, $A_I=\{f,g\}$, $T_E=\{l_1,l_2,l_3,l_4,l_5,l_6,l_7,l_8,l_9\}$, $T_I=\{h_1,h_2\}$.
It holds that $\Sigma(m_0,l_1) = \{h_1\}$ and $\Sigma(m_0,l_5) = \{\varepsilon\}$.
It follows that $\Sigma_{min}(m_0,l_1) = \{h_1\}$, $Y_{min}(m_0,l_1) = \{[1~0]^T\}$, and $\Sigma_{min}(m_0,l_5) = \{\varepsilon\}$, $Y_{min}(m_0,l_5) = \{[0~0]^T\}$.
Let $w=ab$.
It holds that $\hat{\mathcal{J}}(w)=\{(l_1l_2,h1),(l_8l_9,\varepsilon)\}$, and $\hat{Y}_{min}(m_0,w)=\{(l_1l_2,[1~0]^T),(l_8l_9,[0~0]^T)\}$.
The set of basis markings is detailed in Table~\ref{t1}.
The BRG $G$ is shown in Fig.~\ref{fig2}.
\label{example1}
\end{example}

\begin{table}[!t]
\centering
\caption{The set of basis markings.}\label{t1}
\begin{tabular}{cc}
  \hline
  Node & Basis marking\\
  \hline
  $m_0$ & $[1~0~0~0~0~0~0~0~0]^T$ \\
  $m_1$ & $[0~0~1~0~0~0~0~0~0]^T$ \\
  $m_2$ & $[0~1~0~0~0~0~0~0~0]^T$ \\
  $m_3$ & $[0~0~0~0~1~0~0~0~0]^T$ \\
  $m_4$ & $[0~0~0~0~0~0~0~1~0]^T$ \\
  $m_5$ & $[0~0~0~0~0~1~0~0~0]^T$ \\
  $m_6$ & $[0~0~0~0~0~0~1~0~0]^T$ \\
  $m_7$ & $[0~0~0~0~0~0~0~0~1]^T$ \\
  \hline
\end{tabular}
\end{table}

Let $G=(\Theta,E,\Delta,m_0)$ be a BRG and $\mathcal{L}(G)$ be the language generated by $G$.
Let $\sigma = (\lambda_1,y_1)(\lambda_2,y_2)...(\lambda_k,y_k)\in \mathcal{L}(G)$.
We denote by $\varphi: E^*\rightarrow T^*_E$ and $\varphi': E^*\rightarrow \mathbb{N}^{|T_{I}|}$ the sequence and the vector, respectively:
$$\varphi(\sigma) = \lambda_1\lambda_2...\lambda_k$$ and $$\varphi'(\sigma) = \sum\limits_{i=1}^{k} y_i.$$
In other words, $\varphi(\sigma)$ is the concatenation of the first entry in each arc in $\sigma$, and $\varphi'(\sigma)$ is the summation of the second entry in each arc in $\sigma$, namely a j-vector of $\varphi(\sigma)$.
For example, consider the BRG in Fig.~\ref{fig2}, given a sequence $\sigma=(l_1,[1,0])(l_2,[0,0])(l_1,[0,0])$, it holds that $\varphi(\sigma)=l_1l_2l_1$ and $\varphi'(\sigma)=[1,0]+[0,0]+[0,0]=[1,0]$.

\begin{proposition}[\cite{27}]
Let $(N,m_0,l,A)$ be an LPNS and $G=(\Theta,E,\Delta,m_0)$ be the BRG constructed by Algorithm 1.
It holds that
$$\varphi(\mathcal{L}(G)) = P_E(\mathcal{L}(N,m_0)).$$\label{proposition1}
\end{proposition}

In other words, the set of sequences in $\varphi(\mathcal{L}(G))$ coincides with the projection of $\mathcal{L}(N,m_0)$ over the set $T_E$.

\subsection{Unfolded Basis Reachability Graph}

\begin{definition}
Let $(N,m_0,l,A)$ be an LPNS, the \textit{unfolded basis reachability graph (UBRG)} $G_U=(\Theta_U,E_U,\Delta_U,m_0)$ is an NFA, where:
\begin{itemize}
  \item $\Theta_U=\mathcal{M}_b \cup (\mathcal{M}_b \times \{\alpha_i,\beta_j\})$, where $i,j\in\{1,2,3,\ldots\}$;
  \item $E_U\subseteq(T_E \times \mathbb{N}^{|T_{I}|})$ is the set of events;
  \item $\Delta_U\subseteq \Theta_U\times E_U\times \Theta_U$ is the transition relation;
  \item $m_0$ is the initial state.
\end{itemize}
The UBRG can be constructed by Algorithm 2, which also returns two sets ($\Phi_{\alpha}$ and $\Phi_{\beta}$) that are defined later on.\label{def2}
\end{definition}

\begin{algorithm}
\caption{UBRG construction and computation of sets $\Phi_{\alpha}$ and $\Phi_{\beta}$}\label{a2}
\KwIn{An LPNS $(N,m_0,l,A)$, where $A=A_E\cup A_I$.}
\KwOut{The UBRG $G_U=(\Theta_U,E_U,\Delta_U,m_0)$, and two sets $\Phi_\alpha$, $\Phi_\beta$.}

Let $m_0$ be the root node, $\Theta_U = \{m_0\}$, $E_U = \emptyset$, $\Delta_U = \emptyset$, $M_{dup}=\emptyset$.\\
\While{nodes with no tag exist,}
{
  choose a node $m\in\Theta_U$ with no tag,\\
  \If{$m$ is identical to a node in the path from the root node $m_0$ to $m$,}
  {let $M_{dup}=M_{dup}\cup\{m\}$ and goto step 2.}

  \For{all $t\in T_E$,}
  {
    \If{$Y_{min}(m,t)\neq\emptyset$,}
    {
      \For{all $y\in Y_{min}(m,t)$,}
      {
        let $m' = m+[N]_{I}\cdot y+[N](\cdot,t)$,\\
        let $\Theta_U = \Theta_U \cup \{m'\}$, $E_U = E_U\cup \{(t,y)\}$ and $\Delta_U = \Delta_U \cup \{(m,(t,y),m')\}$.
      }
    }
  }
  tag node $m$ ``old''.
}
Let $i=j=0$ and $\Phi_\alpha=\Phi_\beta=\emptyset$.\\
\For{all leaf nodes $m\in\Theta_U$,}
{
    \If{in the path from $m_0$ to $m$ there exists at least one arc whose second entry is not $\overrightarrow{0}$,}
    {
        \uIf{$m\notin M_{dup}$,}{$i=i+1$,\\ replace $m$ with $(m,\alpha_i)$ and $\Phi_\alpha=\Phi_\alpha\cup\{\alpha_i\}$,\\}
        \Else{$j=j+1$,\\ replace $m$ with $(m,\beta_j)$ and $\Phi_\beta=\Phi_\beta\cup\{\beta_j\}$.}
    }

}
Remove all tags.
\end{algorithm}

Hereinafter, we use $m \xrightarrow[G_U]{\sigma} m'$ to denote that $m'$ is reached from $m$ in $G_U$ through a sequence $\sigma$.

\begin{remark}
The UBRG can be viewed as the ``unfolded version'' of the BRG\footnote{Unfolding a graph means transforming a graph - especially one that may have cycles - into a tree-like or acyclic structure that explicitly represents all possible paths starting from an initial node, without looping back. Since we are doing the unfolding of the BRG, we believe that the most appropriate and intuitive name to give to the resulting automaton is unfolded BRG (UBRG), even if it is a tree. This implies that the UBRG may have two identical nodes, which do not have to be merged.}.
The differences between Algorithm 1 and Algorithm 2 are twofold.
The first difference consists in the way repeated nodes are handled: the former fuses repeated nodes, while the latter does not fuse repeated nodes and records a node in the set $M_{dup}$ if it is identical to another node in the path from the root node to the considered one (Steps 4 to 6 in Algorithm 2), where the subscript ``dup'' stands for ``duplicated''.
Therefore, a path of an infinite length in the BRG (i.e., a path ending in a cycle) coincides with a path in the UBRG whose leaf node belongs to $M_{dup}$, and vice versa.
The second difference is that a leaf node (say $m$) in the UBRG may be marked ``$\alpha_i$'' or ``$\beta_j$'' ($i,j\in\{1,2,3,...\}$) if $$m_0\xrightarrow[G_U]{\sigma}m~~\text{and}~~\varphi'(\sigma)\neq\overrightarrow{0},$$
where ``$\alpha_i$'' denotes that $m$ is not a duplicated node (i.e., $m\notin M_{dup}$), while ``$\beta_j$'' denotes that $m$ is a duplicated node (Steps 17 to 28).
In other words, the set of paths in the UBRG whose leaf nodes are marked ``$\alpha_i$'' or ``$\beta_j$'' can identify all the explicit transition sequences whose enabling may only be due to the occurrence of implicit transitions.
The subscripts $i$ and $j$ record the numbers of such two kinds of paths, respectively (Steps 21 and 24).
Finally, $\Phi_\alpha$ and $\Phi_\beta$ denote the sets of ``$\alpha_i$'' and ``$\beta_j$'' in the leaf nodes, respectively (Steps 22 and 25).
\end{remark}

\begin{figure}[!t]
  \centering
  \includegraphics[scale=1]{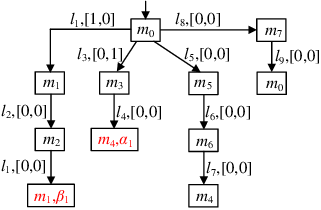}\\
  \caption{UBRG $G_U$ of the LPNS in Fig.~\ref{fig1}.}\label{fig3}
\end{figure}

\begin{example}
Reconsider the LPNS in Fig.~\ref{fig1}, where $T_E=\{l_1,l_2,l_3,l_4,l_5,l_6,l_7,l_8,l_9\}$ and $T_I=\{h_1,h_2\}$.
The UBRG $G_U$ is shown in Fig.~\ref{fig3}, and $M_{dup}=\{m_0,m_1\}$.
We observe that there are two paths in $G_U$ containing implicit transitions, and their leaf nodes are tagged $\alpha_1$ and $\beta_1$, respectively.
It holds that $\Phi_\alpha=\{\alpha_1\}$ and $\Phi_\beta=\{\beta_1\}$.
Indeed, the enabling of both sequences $l_3$ and $l_3l_4$ are due to $h_2$, and the enabling of each sequence in  $\overline{(l_1l_2)^*}\setminus\{\varepsilon\}$ is due to $h_1$, where $\overline{(l_1l_2)^*}$ is the prefix closure of $(l_1l_2)^*$.
\label{example2}
\end{example}

\section{Strong Non-Deterministic Non-Interference Analysis}\label{section4}

In this section, we recall the notion of SNNI and propose a method for the SNNI analysis.

\subsection{The Notion of Strong Non-Deterministic Non-Interference}

In the framework of hierarchical control, a system is observed by users at different levels.
Although all users know the structure of the system, high-level users can observe all the output events (or labels), while low-level users can only observe a subset of output events.
It is possible that there exist some malicious low-level users (e.g., intruders) intending to infer sensitive information.
Therefore, it is necessary to verify if such information leaks may occur under the current hierarchical architecture.

Hereinafter, we denote by $A_L$ the set of labels observable by the low-level users, and by $A_{H}$ the set of labels that may only be observed by high-level users, i.e., $A_{H} = A \setminus A_L$.
Accordingly, $T_{L}$ denotes the set of low-level transitions, i.e.,
$$T_L = \{t\in T\mid l(t) \in A_L\},$$
and $T_{H}$ denotes the set of high-level transitions, i.e.,
$$T_H = \{t\in T\mid l(t) \in A_H\}.$$
We also denote by $[N]_L$ ($[N]_H$) the restriction of $[N]$ to $T_L$ ($T_H$).
The $T_{L}$-induced subnet is denoted by $N_L$.

To use BRG/UBRG for the SNNI analysis, we view the sets $A_L$, $A_H$, $T_L$ and $T_H$ as the sets $A_E$, $A_I$, $T_E$ and $T_I$ in Subsection~\ref{section2.1}, respectively.
Analogously to the projections $P_E$ ($P_I$) defined in Section~\ref{section2}, we define the projections $P_L:T^*\rightarrow T_L^*$ ($P_H:T^*\rightarrow T_H^*$).
Accordingly, the projection of $\mathcal{L}(N,m_0)$ over $T_L$ ($T_H$) is denoted by $P_L(\mathcal{L}(N,m_0))$ ($P_H(\mathcal{L}(N,m_0))$).

Given an LPNS $(N,m_0,l,A)$, the \textit{low-level subnet system} is denoted by $(N_L,m_0,l_L,A_L)$, where $N_L$ is the $T_{L}$-induced subnet, $l_L$ is equal to $l$ restricted to $T_L$.
For example, the low-level subnet system of the LPNS in Fig.~\ref{fig1} can be obtained removing the transitions and the arcs that are highlighted in blue color.

Here we recall the definition of SNNI of LPNSs.

\begin{definition}[\cite{19}]
Let $(N,m_0,l,A)$ be an LPNS and $(N_L,m_0,l_L,A_L)$ be the low-level subnet system.
The PN system is \textit{SNNI} if
$$l(P_L(\mathcal{L}(N,m_0))) = l_L(\mathcal{L}(N_L,m_0)).$$\label{def3}
\end{definition}

In plain words, an LPNS is SNNI if the projection of its language over $T_L$ is indistinguishable from the language of its low-level subnet system.
This implies that low-level users cannot infer the occurrence of high-level transitions after any sequence containing high-level transitions has occurred.

\begin{proposition}
Let $(N,m_0,l,A)$ be an LPNS and $(N_L,m_0,l_L,A_L)$ be the low-level subnet system.
Let $G=(\Theta,E,\Delta,m_0)$ be the BRG constructed using Algorithm~1, where low-level transitions and high-level transitions are viewed as the explicit transitions and the implicit transitions, respectively.
The LPNS is SNNI if and only if
$$l(\varphi(\mathcal{L}(G))) = l_L(\mathcal{L}(N_L,m_0)).$$
\textit{Proof.}~Straightforward from Proposition~\ref{proposition1} and Definition~\ref{def3}.\hfill $\square$
\label{proposition2}
\end{proposition}

This proposition reveals the relation between the language of the BRG and the property of SNNI.

\begin{proposition}
Let $(N,m_0,l,A)$ be an LPNS and $(N_L,m_0,l_L,A_L)$ be the low-level subnet system.
Let $G=(\Theta,E,\Delta,m_0)$ be the BRG constructed using Algorithm~1, where low-level transitions and high-level transitions are viewed as the explicit transitions and the implicit transitions, respectively.
The LPNS is SNNI if and only if
$$\forall s\in\varphi(\mathcal{L}(G)),~\exists s'\in\mathcal{L}(N_L,m_0):~l(s) = l_L(s').$$
\textit{Proof.}~By Proposition~\ref{proposition2}, a system is SNNI if and only if
$$l_L(\mathcal{L}(N_L,m_0)) \subseteq l(\varphi(\mathcal{L}(G)))$$
and
$$l_L(\mathcal{L}(N_L,m_0)) \supseteq l(\varphi(\mathcal{L}(G)))$$
We first observe that
$$l_L(\mathcal{L}(N_L,m_0)) \subseteq l(P_L(\mathcal{L}(N,m_0)))$$
since $(N_L,m_0,l_L,A)$ is the low-level subnet system of $(N,m_0,l,A_L)$.
Therefore, by Proposition~\ref{proposition1}, it always holds that
$$l_L(\mathcal{L}(N_L,m_0)) \subseteq l(\varphi(\mathcal{L}(G))).$$
It means that the LPNS is SNNI if and only if
$$l_L(\mathcal{L}(N_L,m_0)) \supseteq l(\varphi(\mathcal{L}(G))),$$
i.e., for any transition sequence $s\in\varphi(\mathcal{L}(G))$, there exists a sequence $s'\in\mathcal{L}(N_L,m_0)$ such that $l(s) = l_L(s')$.
Hence, the result holds.\hfill $\square$\label{proposition3}
\end{proposition}

By Proposition~\ref{proposition3}, we can analyze SNNI only focusing on the set of sequences in $\varphi(\mathcal{L}(G)$.
Compared with Proposition~\ref{proposition2}, Proposition~\ref{proposition3} provides a more intuitive condition for SNNI.
It should be noted that this proposition may not be directly used for the analysis of SNNI since it is difficult to check all the sequences in $\varphi(\mathcal{L}(G)$ and in $\mathcal{L}(N_L,m_0)$, even if it gives a necessary and sufficient condition for SNNI.
Therefore, in the following subsection, we define a special automaton, called SNNI Verifier (SV), and present a systematic method for SNNI analysis.

\subsection{Strong Non-Deterministic Non-Interference Verifier}

We first give the definition of \textit{parallel composition}, which is inspired by the one in~\cite{34}.
\begin{definition}
Given two NFA with labels $G_1=(\Theta_1,E_1,\Delta_1,m_{1,0},l_L,A_L)$ and $G_2=(\Theta_2,E_2,\Delta_2,m_{2,0},l_L,A_L)$, the \textit{parallel composition} of $G_1$ and $G_2$ is the NFA  $G_1||G_2=(\Theta_1\times\Theta_2,E_1\times E_2,\Delta,(m_{1,0};m_{2,0}))$, where
\begin{itemize}
  \item $((x_1;x_2),(e_1,e_2),(x'_1;x'_2))\in\Delta$ if $(x_1,e_1,x'_1)\in\Delta_1$, $(x_2,e_2,x'_2)\in\Delta_2$, $e_1$ and $e_2$ share the same label;
  \item $((x_1;x_2),(e_1,\varepsilon),(x'_1;x_2))\in\Delta$ if $(x_1,e_1,x'_1)\in\Delta_1$, and the label of $e_1$ is the empty string $\varepsilon$;
  \item $((x_1;x_2),(\varepsilon,e_2),(x_1;x'_2))\in\Delta$ if $(x_2,e_2,x'_2)\in\Delta_2$, and the label of $e_2$ is the empty string $\varepsilon$.
\end{itemize}\label{5}
\end{definition}

Now we introduce the notion of SV.

\begin{definition}
Let $(N,m_0,l,A)$ be an LPNS and $(N_L,m_0,l_L,A_L)$ be the low-level subnet system.
Let $G_U=(\Theta_U,E_U,\Delta_U,m_0)$ be the UBRG constructed using Algorithm 2, where low-level transitions and high-level transitions are viewed as the explicit transitions and the implicit transitions, respectively.
The \textit{SNNI Verifier} $G_V=(\Theta_V,E_V,\Delta_V,v_0)$ is an NFA, where
\begin{itemize}
  \item $\Theta_V=\Theta_U \times R(N_L,m_0)$;
  \item $E_V\subseteq T_L \times T_L$ is the set of events;
  \item $\Delta_V\subseteq \Theta_V\times E_V\times \Theta_V$ is the transition relation;
  \item $v_0\in\Theta_V$ is the initial state.
\end{itemize}
The SV can be constructed by Algorithm 3, which also computes two sets ($\Psi_\alpha$ and $\Psi_\beta$) that are defined later on.\label{def4}
\end{definition}

\begin{algorithm}
\caption{SV construction and computation of sets $\Psi_\alpha$ and $\Psi_\beta$}\label{a4}
\KwIn{An LPNS $(N,m_0,l,A)$.}
\KwOut{The SV $G_V=(\Theta_V,E_V,\Delta_V,v_0)$, and two sets $\Psi_\alpha$, $\Psi_\beta$.}
Construct the UBRG $G_U=(\Theta_U,E_U,\Delta_U,m_0)$ using Algorithm~2, where low-level transitions and high-level transitions are viewed as the explicit transitions and the implicit transitions, respectively.\\
Let $v_0=(m_0;m_0)$, $\Theta_V = \{v_0\}$, $E_V = \emptyset$, $\Delta_V = \emptyset$, $M'_{dup}=\emptyset$.\\
\While{nodes with no tag exist,}
{
  choose a node $(x_1;x_2)\in\Theta_V$ with no tag,\\
  \If{$x_1$ is in the form of $(m,\beta_j)$, and the node $(m;x_2)$ is in the path from the root node $v_0$ to $(x_1;x_2)$,}
  {
  let $M'_{dup}=M'_{dup}\cup(x_1;x_2)$ and goto step 3.
  }
  \For{all $e_1\in E_U$ and all $e_2\in T_L$,}
  {
    \If{$(x_1,e_1,x'_1)\in\Delta_U$, $x_2[e_2\rangle x'_2$ holds in $(N_L,m_0,l_L,A_L)$ and $l(\varphi(e_1))=l_L(e_2)$,}
    {
      $\Theta_V = \Theta_V \cup \{(x'_1;x'_2)\}$, $E_V = E_V \cup (\varphi(e_1),e_2)$, $\Delta_V = \Delta_V \cup \{(x_1;x_2), (\varphi(e_1),e_2)),(x'_1;x'_2)\}$.
    }
  }
  tag node $(x_1;x_2)$ ``old''.
}
Let $\Psi_\alpha=\Psi_\beta=\emptyset$.\\
\For{all the leaf nodes $(x;x')\in\Theta_V$,}
  {
    \If{$x$ is in the form of $(m,\alpha_i)$,}
    {
    $\Psi_\alpha=\Psi_\alpha\cup\{\alpha_i\}$,
    }
    \If{$x$ is in the form of $(m,\beta_j)$, and $(x;x')\in M'_{dup}$,}
    {
    $\Psi_\beta=\Psi_\beta\cup\{\beta_j\}$,
    }
  }
Remove all tags.\\
\end{algorithm}

According to Algorithm~3, the SV $G_V$ is constructed as the parallel composition of $G_U$ and the reachability tree of $(N_L,m_0,l_L,A_L)$, where synchronization is performed on the set $A_L$.
In particular, repeated nodes are not fused and recorded in the set $M'_{dup}$.
In Steps 15 to 23, all the leaf nodes are checked and the information on ``$\alpha_i$'' and ``$\beta_j$'' ($i,j\in\{1,2,3,...\}$) are recorded by the sets $\Psi_\alpha$ and $\Psi_\beta$, respectively.

\begin{figure}[!t]
  \centering
  \includegraphics[scale=1]{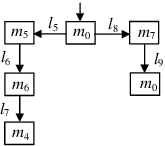}\\
  \caption{The reachability tree of the low-level subnet system of the PN system in Fig.~\ref{fig1}.}\label{fig4}
\end{figure}

\begin{figure}[!t]
  \centering
  \includegraphics[scale=1]{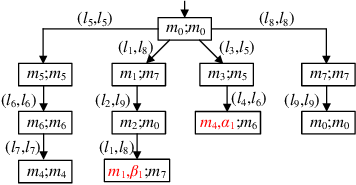}\\
  \caption{SV $G_V$ of the PN system in Fig.~\ref{fig1}.}\label{fig5}
\end{figure}

\begin{example}
Reconsider the LPNS in Fig.~\ref{fig1}.
The reachability tree of its low-level subnet system is shown in Fig.~\ref{fig4}.
The SV is computed using Algorithm~3 and is shown in Fig.~\ref{fig5}.
Note that the transition pair over each arc must share the same label, e.g., $l(l_1)=l(l_8)=a$ and $l(l_3)=l(l_5)=c$.
Furthermore, it is $M'_{dup}=\{(m_0;m_0),(m_1,\beta_1;m_7)\}$, $\Psi_\alpha=\{\alpha_1\}$ and $\Psi_\beta=\{\beta_1\}$.
\label{example3}
\end{example}

\begin{theorem}
Let $G_V=(\Theta_V,E_V,\Delta_V,v_0)$ be the SV constructed using Algorithm~3.
The LPNS is SNNI if and only if
$$\Phi_\alpha=\Psi_\alpha~~\text{and}~~\Phi_\beta=\Psi_\beta,$$
where $\Phi_\alpha$ and $\Phi_\beta$ are computed in Algorithm~2, $\Psi_\alpha$ and $\Psi_\beta$ are computed in Algorithm~3.\\
\textit{Proof.}~Given a sequence $\sigma\in\mathcal{L}(G)$, if $\varphi'(\sigma)=\overrightarrow{0}$, it clearly holds that $\varphi(\sigma)\in\mathcal{L}(N_L,m_0)$.
Therefore, by Proposition~\ref{proposition3}, to analyze SNNI we only need to check the set of sequences in $\mathcal{L}(G)$ whose second entry is not $\overrightarrow{0}$, i.e., the enabling of $\varphi(\sigma)$ may only be due to the occurrence of high-level transitions.
According to Algorithm~2, such sequences can be identified by the set of paths in $G_U$ whose leaf nodes are marked ``$\alpha_i$'' or ``$\beta_j$'', where $i,j\in\{1,2,3,...\}$.
In particular, $\Phi_\alpha$ and $\Phi_\beta$ denote the sets of all ``$\alpha_i$'' and ``$\beta_j$'' in $G_U$, respectively.
According to the two sets $\Psi_\alpha$ and $\Psi_\beta$, one is able to know for which sequences in $\varphi(\mathcal{L}(G_U))$ whose leaf nodes are marked ``$\alpha_i$'' or ``$\beta_j$'', there exist indistinguishable transition sequences in $\mathcal{L}(N_L,m_0)$.
More precisely, by Algorithm~3, $\alpha_i\in\Psi_\alpha$ means that $\varphi(\sigma)$ is of finite length and there exists an indistinguishable low-level transition sequence in $\mathcal{L}(N_L,m_0)$; $\beta_i\in\Psi_\beta$ means that $\varphi(\sigma)$ is of infinite length and there exists an indistinguishable low-level transition sequence of infinite length in $\mathcal{L}(N_L,m_0)$.
Therefore, the equalities $\Phi_\alpha=\Psi_\alpha$ and $\Phi_\beta=\Psi_\beta$ hold if and only if for any low-level transition sequence whose enabling may only be due to the occurrence of high-level transitions, there exists an indistinguishable transition sequence in $\mathcal{L}(N_L,m_0)$.
By Proposition~\ref{proposition3}, the LPNS is SNNI if and only if $\Phi_\alpha=\Psi_\alpha$ and $\Phi_\beta=\Psi_\beta$.
This completes the proof.\hfill $\square$
\label{theorem1}
\end{theorem}

\begin{example}
According to Theorem~\ref{theorem1}, the LPNS in Fig.~\ref{fig1} is SNNI since it holds that $\Phi_\alpha=\Psi_\alpha=\{\alpha_1\}$ and $\Phi_\beta=\Psi_\beta=\{\beta_1\}$.
Indeed, for each transition sequence whose enabling may only be due to the occurrence of high-level transitions, there exists at least one indistinguishable low-level transition sequence in $\mathcal{L}(N_L,m_0)$.
In more detail,
\begin{itemize}
  \item for sequences $l_3$ and $l_3l_4$ in $\varphi(\mathcal{L}(G))$, there exist indistinguishable sequences $l_5$ and $l_5l_6$ in $\mathcal{L}(N_L,m_0)$, respectively;

  \item for each sequence in  $\overline{(l_1l_2)^*}\setminus\{\varepsilon\}\subseteq\varphi(\mathcal{L}(G))$, there exists an indistinguishable sequence in $\overline{(l_8l_9)^*}\setminus\{\varepsilon\}\subseteq\mathcal{L}(N_L,m_0)$.
\end{itemize}
Therefore, low-level users cannot infer the occurrence of high-level transitions.
\label{example4}
\end{example}

\begin{figure}
  \centering
  \includegraphics[scale=0.9]{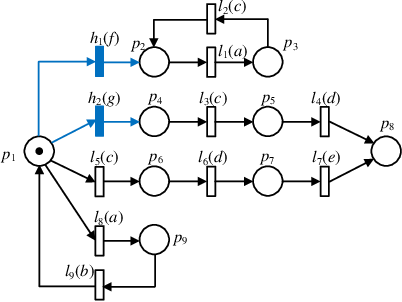}\\
  \caption{A labeled Petri net system.}\label{fig6}
\end{figure}

\begin{figure}
  \centering
  \includegraphics[scale=1]{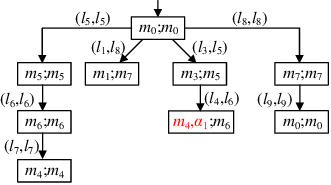}\\
  \caption{SV $G_V$ of the PN system in Fig.~\ref{fig6}.}\label{fig7}
\end{figure}

\begin{example}
Consider the LPNS in Fig.~\ref{fig6}.
Note that the only difference between this LPNS and the one in Fig.~\ref{fig1} consists only in the label of transition $l_2$.
Their basis marking sets and their UBRGs are the same (see Table~\ref{t1} and Fig.~\ref{fig3}).
The SV is computed using Algorithm~3 and is shown in Fig.~\ref{fig7}.
By Theorem~\ref{theorem1}, the LPNS is not SNNI since $\Phi_\alpha=\Psi_\alpha=\{\alpha_1\}$, $\Phi_\beta=\{\beta_1\}$ and $\Psi_\beta=\emptyset$, i.e., $\Phi_\beta\neq\Psi_\beta$.
Indeed, a low-level user may infer the occurrence of $f\in A_H$ after $fac$ has occurred.
\label{example5}
\end{example}

Here we briefly discuss the complexity of the proposed SNNI analysis method.
Let $x$ be the number of reachable markings of the LPNS.
The cardinality of the set of nodes in the BRG and the number of reachable markings of the low-level subnet system are both at most $x$.
It has been proved that the number of basis markings (i.e., the number of nodes in the BRG), in most cases, is much less than the number of reachable markings (see e.g.~\cite{27}).
By Algorithms~2 and~3, the number of nodes in the UBRG and the SV are at most equal to $2x$ and $2x^2$, respectively.
According to Algorithm~3 and Theorem~1, to verify the property of SNNI, we need to check all the leaf nodes in the SV, whose number is at most equal to $2x^2$.
Therefore, the overall complexity of the proposed method is $\mathcal{O}(x^2)$.

\begin{remark}
The problem of SNNI analysis of LPNSs is solved also in~\cite{19}.
To avoid computing the reachability graph, authors in [18] solve some integer linear programming (ILP) problems.
However, the number of constraints may be quite high and cannot be a priori estimated based on the net structure and the number of tokens.
In more detail, for a bounded but non-live system, the number of constraints depends on the dimension of the reachability graph.
For a bounded and live system, such a number depends on the initial marking and on the number of T-invariants.
As a result, the computational efficiency of our method is in general higher than the ILP-based methods.\label{remark2}
\end{remark}

\section{Conclusions}\label{section6}

This paper focuses on the problem of strong non-deterministic non-interference (SNNI) analysis of labeled Petri nets.
We propose a necessary and sufficient condition for SNNI based on a special automaton called SNNI Verifier.
Our future work will be twofold.
First, we will deal with the problem of SNNI enforcement, i.e., design a supervisory control policy to enforce SNNI to a system.
Second, we plan to extend the proposed methods to more properties related to information security of Petri nets such as bisimulation strong non-deterministic non-interference (BSNNI).

\bibliographystyle{unsrtnat}
\bibliography{mybib}
\end{document}